\documentclass[11pt,a4paper]{article}

\usepackage{float}
\usepackage{caption}
\usepackage{subcaption}
\usepackage{algorithm2e}
\usepackage{enumitem}
\usepackage[english]{babel}
\usepackage{hyperref} 
\usepackage[backend=biber,style=numeric,sorting=none]{biblatex}
\usepackage{amssymb}
\usepackage{amsmath}
\usepackage{amsthm}
\usepackage{graphicx} 
\usepackage{hyperref} 
\usepackage{perpage} 
\MakePerPage{footnote} 
\usepackage[margin=2cm]{geometry} 
\fontencoding{T1} 
\usepackage{chngcntr} 
\usepackage{import}
\usepackage{pdfpages}
\usepackage{transparent}
\usepackage{xcolor}

\usepackage{listings}
\definecolor{codegreen}{rgb}{0,0.6,0}
\definecolor{codegray}{rgb}{0.5,0.5,0.5}
\definecolor{codepurple}{rgb}{0.58,0,0.82}
\definecolor{backcolour}{rgb}{0.95,0.95,0.92}

\lstdefinestyle{mystyle}{
    backgroundcolor=\color{backcolour},   
    commentstyle=\color{codegreen},
    keywordstyle=\color{magenta},
    numberstyle=\tiny\color{codegray},
    stringstyle=\color{codepurple},
    basicstyle=\ttfamily\footnotesize,
    breakatwhitespace=false,         
    breaklines=true,                 
    captionpos=b,                    
    keepspaces=true,                 
    numbers=left,                    
    numbersep=5pt,                  
    showspaces=false,                
    showstringspaces=false,
    showtabs=false,                  
    tabsize=2
}
\counterwithin{figure}{section}

\SetCommentSty{mycommfont}

\theoremstyle{plain}

\theoremstyle{definition}

\theoremstyle{remark}

\title{A Survey on Data Cleaning Methods for Improved Machine Learning Model Performance}
\author{Ga Young Lee, Lubna Alzamil, Bakhtiyar Doskenov, Arash Termehchy}
\date{April 13, 2021}

\begin{document}
\maketitle
\section{Abstract}
Data cleaning is the initial stage of any machine learning project and is one of the most critical processes in data analysis. It is a critical step in ensuring that the dataset is devoid of incorrect or erroneous data.
It can be done manually with data wrangling tools, or it can be completed automatically with a computer program. Data cleaning entails a slew of procedures that, once done, make the data ready for analysis. Given its significance in numerous fields, there is a growing interest in the development of efficient and effective data cleaning frameworks. In this survey, some of the most recent advancements of data cleaning approaches are examined for their effectiveness and the future research directions are suggested to close the gap in each of the methods\cite{1}.

\section{Introduction}
The availability of high-quality data is typically a prerequisite for developing high performing accurate Machine Learning (ML) applications.
However, data is rarely clean in reality due to noisy inputs from manual data curation or inevitable flaws from automatic data collection or generation processes. For example, the inconsistency and incompleteness caused by broken sensors or human-error are common in real-world datasets, and can impact the machine learning systems built on top of them. To address this problem, we reviewed and propose several state-of-the-art data cleaning approaches based on literature review of publications in the data management systems field. 
\subsection{Problem Statement}

While some tools in the market aim to solve relatively easy problems, such as imputing missing values, there still exists a great demand for solving more complex issues such as inconsistent notations of the same information. A majority of the traditional data cleaning frameworks require the input from users to identify irregularities and remove the erroneous values applying any update, which poses a huge burden in terms of time and labor cost. Additionally, the sequence of various data cleaning methods determines the outcome of data cleaning, so extra consideration on the relations between them should be made to ensure the proper sequencing of data cleaning techniques. \newline{}

Given this condition, the solution space of combining data cleaning methods can be enormous, where each combination represents a pipeline with the possibility to define various parameters. Since the sequence and the parameters are predefined, this leads to several issues, which is the lack of parameter tuning as cleaning progresses through the pipeline as well as a growing need for reordering cleaning scripts to find a better combination\cite{2}. These issues remain an important question to answer for many researchers in the database managment systems field. Some of the notable researchers including Sanjay Krishnan and Eugene Wu propose evolving solutions to address the limitations of the previous approaches described in section 3 Relevant Works.

\subsection{The Negative Impact of Dirty Data}

Data cleaning to ensure the consistency of training data is an essential step for maintaining the model performance because inconsistencies and errors present in the training data can detract algorithms from detecting patterns. However, this process takes multiple iterations to reach the equilibrium point where the original data is sufficiently cleaned to represent the accurate distribution independent of potential biases and errors. The consequence of poorly handled data can lead to wasted resources, lost productivity, failing internal and external communication, and wasted marketing investment. According to a market research, it is estimated that erroneous or incomplete customer and prospect data wastes 27\% of revenue. As a result, dirty data has been a critical bottleneck in increasing productivity and managing resources efficiently in several domains, such as finance, electronic health records, and e-commerce\cite{3}.

\subsection{The Importance of Data Cleaning}
According to a survey conducted by Forbes, up to 60\% of data scientists' time is spent on cleaning, standardizing, and organizing data\cite{4}.
Meanwhile, knowledge workers spend up to 50\% of their time dealing with obfuscated and incorrect data\cite{5}. Because dirty data lacks credibility, end-users who rely on it must spend more time verifying its accuracy, lowering speed and productivity even more.
Adding additional manual process leads to increased inaccuracies and inconsistencies as the number of dirty records grows. Aside from the financial loss, stale data has a more subtle influence on firms\cite{3}. Therefore, understanding the impact of such inconsistencies is essential in that they help in understanding various practical implications such as; a large enough number of incomplete information does not have an impact on ML models, thus spending so much effort on trying to clean or acquire specific data or a piece of missing information does not change the quality of downstream Machine learning models. Data cleaning is, in most cases, an essential prerequisite in the development of an ML application. However, most research considers data cleaning as a standalone exercise regardless of its impact on ML applications. Data cleaning can degrade the performance of ML models\cite{6}.

\section{Relevant Works}

Considering the importance of data cleaning in the downstream machine learning projects, numerous researchers have made progress in developing novel solutions for data cleaning. In particular, some of the recent works published after 2015 are examined and compared for their strengths and weaknesses. The state-of-the-art data cleaning approaches covered in the survey paper are SampleClean\cite{7}, ActiveClean\cite{8}, Holistic Data Cleaning\cite{9}, AlphaClean\cite{2}, and CPClean\cite{6}. These methods are closely reviewed in the following subsections respectively.

\subsection{SampleClean: Simulated Clean Data Instances}
SampleClean suggests a solution to sample the raw data that can better present clean data instances. Naive sample approach can be misleading because the semantics of dirty data is different from the semantics of clean data, so what the researchers used to solve this problem is called Approximate Query Processing (AQP)\cite{7}. The AQP consists of two steps: first, in Direct Estimate (DE), a set of k rows is sampled randomly and cleaned, and the training result is returned independently of the dirty data.  However, DE alone can lead to wrong answers when the impact of dirty data is dominant in the sampled subset. To mitigate this drawback, Correction step is used to reweight the sample based on the contribution of the cleaned data to the whole dataset when it’s used in training. By adopting this framework, Krishnan et al were able to reach the equilibrium where the approximate query result is bounded by the confidence interval that can ensure the proper amount of clean data is captured in the simulation. To do so, a probabilistic model was adopted to reweight samples based on their contribution to the whole dataset. In addition, a soft boundary was imposed to ensure the proper amount of data is captured in the simulated sample. For example, if a user wants to query data to get a certain amount of accurate data, SampleClean suggests the sample size recommended to meet the desired level of accuracy bounded asymptotically by estimators. Also, by leveraging “Stochastic Gradient Descent,” the average value of transformed data is calculated and converges at the optimal value. However, the simulation of cleaned data instances insufficient for identifying the correct version of clean data. To address this issue, the same group of researchers published a follow-up study on the data cleaning method called ActiveClean. 

\subsection{ActiveClean: Incremental Data Cleaning in Convex Models}
Unlike SampleClean which suggests constructing a simulated clean data instance, the researchers suggested iteration retaining the models based on the optimizers. The optimizers are “Stochastic Gradient Descent” that iteratively samples data, estimates a gradient and updates the current best model. By applying this method, practitioners only need to a small amount of data as opposed to the complete retraining. The objective of ActiveClean is to learn a model over dirty data without cleaning and transforming it. ActiveClean gradually cleans a dirty dataset to learn a
convex-loss model, such as Logistic Regression and Support Vector Machine (SVM). The key difference is that ActiveClean aims to clean the underlying dataset incrementally such that the learned model becomes more effective as it receives more cleaned records. However, this approach has some drawbacks. For example, the applicability of ActiveClean is limited to Convex models, and it requires user input such as the specification of the model, gradient, and stopping criteria.

\subsection{HoloClean: Holistic Data Repairs With Probabilistic Inference}
The improvements in probabilistic inference helped to create another framework for data cleaning\cite{9}. The main difference compared to other approaches in the field is that it considers marginal probability when reasoning about the candidate sample to clean. The statistics and the dataset provided by customers helped the model to reason about the nature of the "dirty" records. As an initial approach, the authors developed a model, which learns on both clean and corrupted data to infer where potential errors are highly likely to occur. The authors propose training a model on a relatively small portion of real data and checking if it can infer the nature of the mistakes done by humans when filling out the data. Next, when the authors tested the model on a whole dataset, the test accuracy results were more than 90\%, which showed that the proposed inference model is promising. To train an optimal model, HoloClean needs both clean and dirty samples to train a model. This might require cleaning and verifying the training set. However, in the reality, it is hard to get a sufficient amount of clean and dirty data samples to learn on.

\subsection{AlphaClean: Generate-Then-Search Parallel Data Clenaing}
The authors of AlphaClean\cite{2} propose a system, which would automate the process of finding an effective pipeline of cleaning tools. The process of building a candidate pipeline includes parameter selection for the specific, generated pipeline. The AlphaClean solution works with any given amount of cleaning tools, which potential users work with to clean the data. The number of given tools should be reasonably limited due to the way the algorithm generates the cleaning pipeline. It builds a combination of pipelines in a tree-shaped manner. As the algorithm progresses, the solution gets available progressively and the output is revised in further iterations, which is supposed to improve the quality of the proposed pipeline. Those pipelines can be accessed and tested by the user at any given time when the algorithm runs. This frameworks allows the flexibility to assess the outcome from each step, where users can check if the proposed branch of data cleaning pipelines is most promising and should be considered as a candidate solution.

\subsection{CPClean: Reusable Computation in Data Cleaning}

In the Nearest Neighbor Classifiers over Incomplete Information paper\cite{6}, The authors proposed several solutions to curb the problem of these inconsistencies that are assumed to impact machine learning. Such solutions include using checking and counting as tools for studying the impact of incomplete data on training machine learning applications. The paper also proposes an extensive data collection for the machine learning approach, the CPClean, which is developed on top of the CP primitives whose performance has more significance on existing work, especially on datasets with systematic values. Such is seen when comparing conventional data collecting approaches with CPClean. When working on 5 datasets, all with systematic missingness, CPClean closes a 100\% gap on average by cleaning up to 36\% of dirty data. In contrast, the other data collecting approaches, such as BoostClean, which is the best automatic cleaning approach, can only clean a 14\% gap on average. CPClean has an advantage over other cleaning approaches because it is most appealing on data sets with systematic missingness as it closes 100\% in all cases where other techniques such as BoostClean fail to achieve good performance\cite{6}.
	


\section{Open Problem}
\subsection{Lack of Optimizer} 

The current data cleaning solutions lack in the clear interface that suggests the optimal stopping point of the amount of data to be cleaned. As a result, practitioners rarely reach the equilibrium required to ensure the desirable level of clean data in the training set. This problem is briefly discussed in ActiveClean where users can set the parameters such as models, gradient, and stopping criteria. However, these manual manipulations require the domain expertise and a deep understanding of the parameter tuning method. Therefore, the applicability of ActiveClean's optimizer is limited to a small subset of users who can benefit from data cleaning. In other words, the limitation in the "ActiveClean" model is that the user must be a technician to open the model and check why the machine gives poor predictions. Opening the diagnose panel is challenging for users who do not know about machine functionalities. Further, when the user cleans the data by mistake or cleans, they need the data; the model does not show where the cleaned data is stored or if it will never be recovered. Therefore, further direction for the model is that it should create a panel for cleaning through removing and correcting the data depending on how the user needs it. To overcome this challenge, data checking models can be designed depending on data problems and how users prefer. Therefore, it is suggested a future data cleaning solution should consider users' needs and problems to be solved and communicate its progress with them through the process.

\subsection{Tradeoff Between Efficiency and Coverage}
The earlier data cleaning techniques, such as SampleClean and HoloClean, focus on the extensive data cleaning that covers the entire data instances, which can guarantee a high level of coverage at a cost of expensive computing resources. On the other hand, the later developments using inference and logic show a promise of using as little data as possible to generate clean instances. However, due to the nature of the sparsity observed in raw data, this approach of using inference and learning fails to ensure the optimal coverage of the data. This tradeoff between efficiency and coverage remains a perennial problem in data cleaning to address by designing an efficient algorithm or using cache to store the preprocessed subset of raw data. It is suggested in the research that with the advent of widely available high-performing computing resources, data cleaning moves toward the direction of less cost with more coverage. The drawback of the iterative approach is discovered in AlphaClean: AlphaClean registers complex systems due to its numerous simultaneous functionalities. It is challenging to generate cleaning language for various systems due to different modes and systems of data productions. The repair functions are complex, but sometimes they do not fulfill their functionalities. Despite the parallelism being a beneficial tool for enhancing performance, the systems is prone to repair each instance in most of the times. Also, slow cleaning operators pose a significant burden on overall speed. Similarly, such systems may result in slow quality evaluation systems. The sequential data cleaning process cannot perform heavy functionalities in different databases. Therefore, the process systems are not effective for a large-scale database solution. In AlphaClean\cite{2}, the authors suggest that the future work can be done to make it flexible and integrate AlphaClean with other visualization systems, making it easy and allowed for quality functions.   

\subsection{Limited Generalizability} 
The empirical results from the data cleaning methods indicate that some methods are hyperspecific to a certain type of data. In order to expand the scope of the data cleaning ability, it is crucial to design and test data cleaning methods that can adapt to a new given dataset and learn from. This problem is most notable in CPClean. Specifically, the limitation of the Nearest Neighbor Classifiers over Incomplete Information paper\cite{6} is that they did their experiments on just five datasets. This number is arbitrary and hard to justify for other applications. There is a high likelihood that the method might not work well with different experiment setting. Another limitation is that it is not clear whether their approach will work with other classifiers or just the K-Nearest Neighbor Classifier. Their approach with K-Nearest Neighbor gave linear complexity which is great. However, they did not provide any reasoning behind the choice of KNN beyond the simplicity of the algorithm. The future work for this paper would be to test with other classifiers and see if it possible to reach similar complexity.

\section{Future Research Directions} 

\subsection{Data Visualization for Optimization}
Data visualization can be integrated into data cleaning so that users can keep track of the progress made with data cleaning and visually manipulate data. Specifically, visual data cleaning can be translated into quality functions; users can use data visualization to study data that would otherwise be merely a collection of tables, figures, and text. Texts containing images have been demonstrated to be easier to comprehend in several studies. As a result, data visualization communicates the area of improvement by combining figures and tables with visual features. It displays the interaction between two or more aspects and shows multiple points of one particular aspect, for example, by doing so. It's a means to improve the viewer's experience and make data cleaning more useful\cite{10}. 

\subsection{Combining Data Programming With Data Cleaning}
Exploring how data programming and data cleaning can be unified under a common probabilistic framework to perform better detection and repairing. Designing a framework that integrates programming with data cleaning will help keep a closer eye on your errors to assist you delete data that is erroneous, corrupt, or inconsistent. An expected benefit of this approach is that uses will make fewer mistakes. Additionally, different functions and what your data should do can be mapped in the interface. Errors in many data sources can be easily removed by taking this approach\cite{11}.

\subsection{Exploring Hardware Solutions for Memory Management}
Memory management is a type of operating system management system whose purpose is to organize and manage data in a systematic manner. The following are some of the benefits: a user can get or place raw data to a specific area, data processing capability improves, the amount of time spent loading is minimized, and the management of RAM improves and Data security improves\cite{11}. Trying to Lower the cost of cleaning the data and increase the coverage of the datasets can be enhanced by exploring aforementioned hardware solutions. Considering the enormous computation power required in a large-scale database management system, the GPU and ASIC capabilities can suggests a breakthrough in data cleaning by efficient memory management. In particular, the ability to perform the computation in parallel or to be portable to other hardware architectures can be beneficial for the recent advancements in the large-scale database management solutions involving data cleaning\cite{2}.

\section{Summary}
Due to various reasons such as broken sensors or human errors, the result of data collection can be misleading and contain a number of quality issues such that the information can be missing at random or misrepresented by inaccurate notations. The consequence of dirty data is that it can undermine the quality of machine learning projects in the downstream workflow and can even cause a financial cost and damage trust. When integrated properly, data cleaning can mitigate such problems. Therefore, data cleaning is a necessary step to consider in any data project and should be invested in to guarantee accurate and reliable information. To that end, this survey paper presents several approaches to data cleaning developed by reputable researchers and developers in the database management systems field to solve the challenge with data cleaning. Various data cleaning techniques are analyzed and compared; a general trend observed from the literature review is a transition toward more scalable and efficient frameworks that aims to reduce the human overhead cost in the development of more accurate and representative data instances. There are several areas of improvement such as integrating with visual interface and usage of high-performance memory management hardware solutions. With the current rate of progress, a novel data cleaning solution will alleviate the listed problems and make an impact in numerous fields by guaranteeing the data quality while communicating the efficient data cleaning workflow with users.

\printbibliography 
\end{document}